\documentclass[fleqn,usenatbib]{mnras}
\usepackage{mathptmx}
\usepackage{txfonts}
\usepackage{float}
\usepackage[T1]{fontenc}
\usepackage{ae,aecompl}
\usepackage{graphicx}	
\usepackage{hyperref}
\usepackage{amssymb}	
\usepackage[dvipsnames]{xcolor}
\hypersetup{colorlinks=true,linkcolor=black,citecolor=black,filecolor=black,urlcolor=black}

\title{Periodicities in the {\textit K2} lightcurve of HP Librae}
\author[S. Solanki et al.]{
Siddhant Solanki,$^{1}$\thanks{E-mail: siddhantsolanki@ucsb.edu }
Thomas Kupfer,$^{2,3}$
Omer Blaes,$^{1}$
Elm\'e Breedt,$^{4}$ and
\newauthor Simone Scaringi$^{3,5}$
\\
$^{1}$Physics Department, University of California, Santa Barbara, CA 93106, USA\\
$^{2}$Kavli Institute for Theoretical Physics, University of California, Santa Barbara, CA 93106, USA \\
$^{3}$Department of Physics and Astronomy, Texas Tech University, PO Box 41051, Lubbock, TX 79409, USA\\
$^{4}$Institute of Astronomy, University of Cambridge, Madingley Road, Cambridge, CB3 0HA, United Kingdom \\
$^{5}$Centre for Extragalactic Astronomy, Department of Physics, University of Durham, South Road, Durham, DH1 3LE, UK\\
}

\date{Accepted XXX. Received YYY; in original form ZZZ}
\pubyear{2020}

\begin{document}
\label{firstpage}
\pagerange{\pageref{firstpage}--\pageref{lastpage}}
\maketitle

\begin{abstract}
    We analyse \textit{Kepler/K2} lightcurve data of the AM CVn system HP Librae.  We detect with confidence four photometric periodicities in the system: the orbital frequency, both positive and negative superhumps, and the positive apsidal precession frequency of the accretion disc.  This is only the second time that the apsidal precession frequency has ever been directly detected in the photometry of a helium accreting system, after SDSS J135154.46-064309.0.  We present phase-folded lightcurves and sliding power spectra of each of the four periodicities.  We measure rates of change of the positive superhump period of $\sim10^{-7}$~days/day.  We also redetect a QPO at $\sim300$~cyc/day, a feature that has been stable over decades, and show that it is harmonically related to two other QPOs, the lowest of which is centered on the superhump/orbital frequency.  The continuum power spectrum is consistent with a single power law with no evidence of any breaks within our observed frequency range.  
\end{abstract}


\section{Introduction}
    HP\,Librae (HP\,Lib) is one of the brightest members of the class of AM\,CVn systems, a group of ultra-compact mass transferring binaries consisting of a white dwarf primary and a degenerate or semi-degenerate secondary. Observed orbital periods in these systems are extremely short:  from 5.4 min (HM Cnc, \citealt{roe10}) to 67.8 min (SDSS~J1505, \citealt{gre20}).  As a result, AM\,CVn systems are predicted to be strong, low-frequency, galactic gravitational wave sources \citep{nel04, roe07c, nis12, kre17, kup18}.  They may also be the source population of proposed .Ia supernovae \citep{bil07}, and are in general useful probes of the final stages of binary evolution. Spectroscopically they are characterised by strong helium lines and a complete lack of hydrogen, which is an indicator of the advanced evolutionary state of the systems. 
    
    AM\,CVn systems are thought to start mass transfer from a helium rich donor star at orbital periods $\leq\,15$\,minutes and evolve to longer periods. They likely span a very large range of accretion rates $(\approx10^{-5}$-$10^{-12}$\,M$_\odot$\,yr$^{-1})$ over the course of this evolution \citep{del07}. As the orbital period increases and the accretion rate decreases, AM\,CVn systems pass through distinct phenomenological states. The most compact systems with orbital periods $\leq\,10$\,min do not show signs of an accretion disc and are most likely in a state where the accretion stream hits the white dwarf primary directly \citep{mar02,esp05,roe10}. For orbital periods between $\approx$10\,min and $\approx$20\,min, these systems are observed in a constant high state where an optically thick accretion disc with helium absorption lines dominates the observed optical flux. HP\,Lib is a member of this class of high-state AM\,CVn systems, and other members include AM\,CVn itself, SDSS\,J1908, CXOGBS\,J1751, and SDSS\,J1351 \citep{fon11,kup15,wev16,gre18a}. At the longest orbital periods $\geq\,40$-$50$\,min, AM\,CVn systems have low accretion rates and are in a permanent low state with an optically thin accretion disc exhibiting helium emission lines (e.g. \citealt{mor03}). Systems with intermediate periods undergo outbursts between low and high states, which can be modeled as a helium ionization version of dwarf nova outbursts (e.g. \citealt{sma83,can84,kot12,can15}). The outburst frequency decreases towards longer periods while the outburst amplitudes increase with longer periods \citep{lev15}. For recent reviews of AM\,CVn systems, see \citet{nel05}, \citet{sol10} and \citet{bre15}.
    
   Photometric power spectra of high-state AM\,CVn systems and systems in outburst often show multiple peaks that can be related to harmonics and sums and differences of at least three distinct types of variability (e.g. \citealt{ski99}).
   One is the orbital period which is generally confirmed by spectroscopy (e.g. \citealt{kup15}). The other two are positive (longer than the orbital period) and negative (shorter than the orbital period) superhumps.  The positive superhumps are the most commonly observed, and are widely believed to be due to prograde apsidal precession of an eccentric disk \citep{whi88}, as in SU UMa systems.  Negative superhumps may be due to retrograde nodal precession of a tilted disk, though a physical explanation as to how both types of superhump can be simultaneously present in the same system is as yet unclear.  The difference in orbital period and the superhump periods should correspond to the relevant precession period, which for the positive superhump has been observed directly in spectroscopic variability of AM~CVn itself \citep{pat93} and recently in photometric variability of SDSS~J1351 \citep{gre18a}. 
   
    
    \begin{figure*}
        \centering
        \includegraphics[scale=0.4]{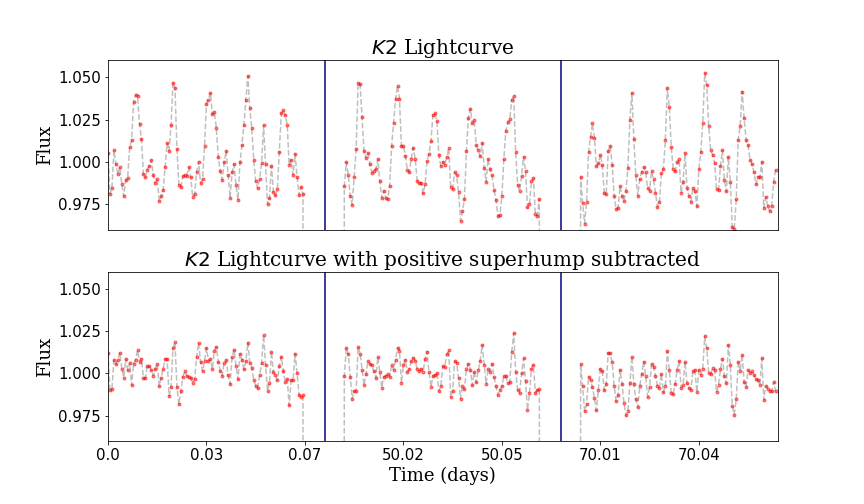}
        \caption{Sections of the K2 lightcurve taken at 3 different times of 
        observation before (top) and after (bottom) subtraction of the positive superhump. The variations between successive superhump cycles in the
        top panel are more than the amplitude of the orbital period, and as a result the superhump might 
        still have significant residual contributions even after subtraction. A movie of the entire 88 day 
        lightcurve can be found in the online supplementary materials
        for this paper.
        } 
        \centering
        \label{fig:hp-lib-lc}
    \end{figure*}

        \begin{figure*}
        \centering
        \includegraphics[scale=0.35]{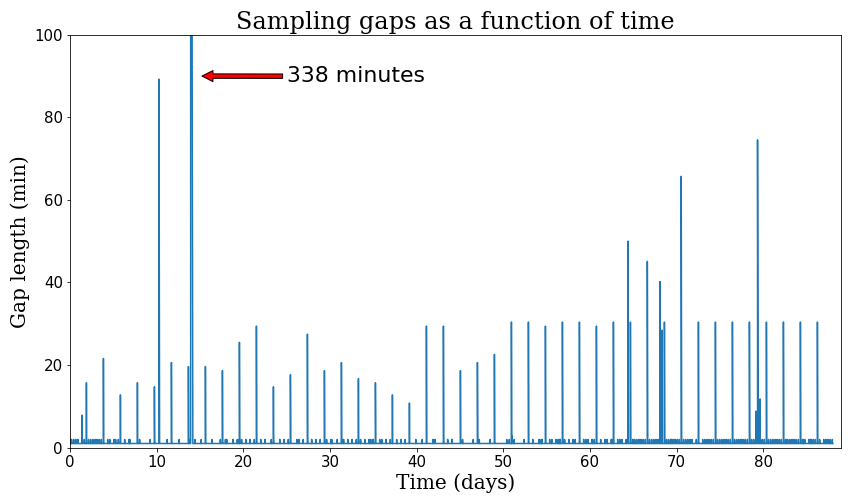}
        \caption{Data gaps in the light curve over the {\it K2} observation. Note that some of the gaps have an approximate two-day periodicity.  The vast majority of the observations are separated by a one minute cadence. There is one 
        gap which is 338 minutes long, as annotated in the figure.}
        
        \centering
        \label{fig:lc_properties}
    \end{figure*}
    
    HP Lib was first discovered by \citet{odo94} in the Edinburgh-Cape blue object survey as only the sixth known AM\,CVn star. They found rapid oscillations at a 1119\,sec period as well as broad, shallow He\,{\sc i} absorption lines in the spectrum, a characteristic feature of these high state systems. \citet{pat02} performed a photometric variability campaign over 6 years and found a dominant periodicity varying between 1118.89 and 1119.14\,s in that time-frame. In addition to this primary variation, \citet{pat02} detected a much lower amplitude modulation of 0.005\,mag with a period of 1102.70$\pm$0.05\,sec. The dominant variation at 1119\,sec was interpreted as the positive superhump, whereas the low amplitude modulation was interpreted as the orbital period. This was confirmed by \citet{roe07} who performed phase resolved spectroscopy of HP\,Lib and found an orbital period of 1102.8$\pm$0.2\,sec. With its close distance ($276\pm4$~pc, \citealt{ram18}) and such a short period, HP\,Lib will be among the strongest known sources of gravitational-wave emission for the \textit{Laser Interferometer Space Antenna} (LISA; \citealt{ama17, kup18}).  
    
   
   Studies using space-based photometry for AM\,CVn systems are sparse but can produce a rich data set. \citet{fon11} discovered SDSS\,J1908 as a new high-state AM\,CVn system in the original field observed by the \textit{Kepler} space observatory mission.  A follow-up study using the full 4-year \textit{Kepler} data set revealed 42 independent frequencies and an orbital period very similar to HP\,Lib \citep{kup15}. A new high-state system was discovered in the extended \textit{Kepler} mission (\textit{K2}) in campaign 6 \citep{gre18a}.  HP\,Lib was observed in campaign 15 of \textit{K2} and was included for short cadence observations, providing nearly continuous photometry for 88\,days with a 1\,min cadence.  Here we present a detailed analysis of the \textit{K2} lightcurve.


\section{Observations and Data Analysis}\label{sec:methods}

     The original pixel data were downloaded from the Kepler Data Archive, covering a total baseline of 88 days. The lightcurve was extracted using aperture photometry and was normalised and flat-fielded using the PyKE tools provided by the NASA Kepler Guest Observer Office \citep{sti12}. Depicting the entire resulting 88 day lightcurve in a single figure is not very informative due to the rapid superhump variations, but we show three representative epochs in the top panel of  Fig. \ref{fig:hp-lib-lc} \footnote{A movie showing the entire lightcurve can be found in the online supplementary materials for this paper.}.  (Fig. \ref{fig:binnedlightcurve} below also shows the entire lightcurve binned on 250 minute intervals.)
    There were gaps in the data set, possibly due to the telescope maneuvers, and a histogram of these gaps is shown in  Fig.~\ref{fig:lc_properties}.
    Because the vast majority of the observations had continuous one minute sampling, we safely took the Nyquist frequency to be twice per minute or 720 cyc/day.  However, the approximately two-day gap periodicity that is evident in this figure means that one cannot trust the power spectrum below a frequency of $\simeq0.5$~cyc/day. \newline

    The periodogram shown in Fig.~\ref{fig:lombscargle} below
    was calculated using Astropy's {\tt Lomb-Scargle} routine \citep{Lom76, sca82}.\footnote{https://docs.astropy.org/en/stable/timeseries/lombscargle.html}  The frequency limits on the periodogram were set from $\frac{1}{44}\simeq0.02$~cyc/day to the Nyquist frequency of 720 cyc/day.  To judge the significance of the peaks in the periodogram, a line of height 4.5 times the rms over 3 day\(^{-1}\) bins was plotted over the periodogram. Peaks that exceeded a certain power threshold (\(1 \times 10^{-4} \) units in power) in the periodogram were not included in the rms calculation. The 4.5 sigma threshold has been chosen as a balance between recovering the most signals while minimising the excessive false positive rate caused by the underlying noise process (e.g. \citealt{tel12, kup15, ree18}).
    
    In order to construct sliding periodograms, the positive superhump had to be subtracted from the lightcurve. This was because it contributed a changing flux that was over an order of magnitude larger than the variability at other frequencies. The subtraction was done by breaking the lightcurve into 1 day intervals.  {\tt Lomb-Scargle} was then run to calculate the superhump frequency, $\omega_{\rm sh}$, in each interval.  Then the function $A(\sin\omega_{\rm sh}t)+B\cos(\omega_{\rm sh}t)$ was fitted to each mean subtracted light curve in that interval, where the parameters $A$ and $B$ were found through scipy's {\tt curve\_fit} routine.\footnote{https://docs.scipy.org/doc/scipy/reference/generated/
    scipy.optimize.curve\_fit.html} This fit was subtracted off the lightcurve, and then the process was repeated successively at the superhump harmonics until the orbital period was the dominant feature in the power spectrum. The lower panel of Fig. \ref{fig:hp-lib-lc} shows the superhump subtracted lightcurve that we obtained. The superhump was not cleanly subtracted because of significant variations in the successive cycles of the superhump. The different panels in the figure were chosen at times when the negative superhump signal was strong (panels 1 and 3), and when the negative superhump signal was weak (panel 2). There is little apparent difference by eye.  After subtraction of the positive superhump, we were then able to construct sliding periodograms at the other frequencies (see Fig.~\ref{fig:sliding_fft} below).
    
\section{Results}

Figs.~\ref{fig:lombscargle}-\ref{fig:lombscargle_sh} show our Lomb-Scargle power spectrum for the entire {\it K2} lightcurve.
For detection of discrete features in the power spectrum, we adopt a threshold of $4.5\sigma$ above the local rms as described in Sec.\,\ref{sec:methods}. We find a total of 18 significant frequencies in the Lomb-Scargle periodogram, all listed in Table~\ref{tab:frequencies}.  All 18 result from harmonics and/or combinations of four physically fundamental signals at 77.205, 78.354, 1.15 and 79.3 cyc/day, in order of decreasing strength.  We identify these as the positive superhump, orbital frequency, apsidal precession frequency, and a small feature which might be a negative superhump. The positive superhump frequency agrees with the dominant periodicity detected by \citet{pat02} and \citet{odo94}, and the orbital frequency feature, also detected by \citet{pat02}, agrees with the spectroscopic orbital frequency measured by \citet{roe07}. The positive superhump frequency is consistent with being the difference between the orbital and apsidal precession frequencies.  Note that while the putative negative superhump is the weakest of the four fundamental frequencies, it is noteworthy that we also detect one intermodulation frequency that is related to it, increasing its significance still further.  In addition to the sharp frequency features, we also detect a quasi-periodic-oscillation (QPO) at a frequency of about 250–350 c/d corresponding to a period of 4–6 min, again confirming a previous detection by \citet{pat02}.   We will briefly discuss this QPO further at the end of this section where we will show that there are two additional QPOs to which it is harmonically related.  (All these QPOs are more clearly visible on a log-log periodogram, as shown in Fig. \ref{fig:power-law} below.)
    
    Fig.~\ref{fig:phase_fold} shows the phase-folded lightcurves of the four fundamental frequencies. The positive superhump is the strongest, with an amplitude more than 10 times larger than the next strongest variations which are the orbital period and precession period. The positive superhump is directly visible in the phase-folded {\textit K2} lightcurve, whereas for the other frequencies we removed the signal of the positive superhump from the light curve before phase-folding, as described in Sec.\,\ref{sec:methods}. The phase folded lightcurves on the orbital and precession periods show comparable amplitudes in addition to having nearly equal peak heights at their respective frequencies in the periodogram. This might suggest that the radiating areas responsible for the two frequencies are similar.\footnote{We thank the referee for pointing this out to us.}
    
   As illustrated in the top panel of Fig.~\ref{fig:hp-lib-lc}, the shape of the positive superhump varies significantly from period to period.  Nevertheless, the positive superhump generally shows two maxima, where one maximum is about three times stronger than the other maximum. The mean phase-folded light curve folded on the orbital period is flatter at maximum and sharper at minimum.  The shape of the mean phase-folded light curve folded on the precession period is opposite, with the maximum being sharper.
    

    \begin{table*}
     
     \label{tab:freq}
     \begin{tabular}{||c | c | c | c ||}
      \hline
      Type & Frequency (cyc/day) & Period (sec) & Relation\\
      \hline \hline
      Positive Superhump & 77.20544 \(\pm\) 0.00004 & 1119.1 & \(\omega - \Omega\)\\
      \hline
      Orbital & 78.354 \(\pm\) 0.002 & 1102.7 & \(\omega\) \\
        \hline
      Apsidal Precession & 1.1501 \(\pm\) 0.0002 & 75214.9 &\(\Omega\)\\
        \hline
      Negative Superhump (?)  & 79.3 \(\pm\) 0.01 & 1089.5 &\(\omega + N \)\\
      \hline
      \(2^{nd}\) SH Harmonic  & 154.41014 \(\pm\) 0.00004 & 559.5 & $2\omega - 2\Omega$\\
      \hline
      ?? & 155.5601 \(\pm\)  0.0004 & 555.4 & $2\omega - \Omega$\\
      \hline
      ?? & 156.528 \(\pm\) 0.001 &  552.0 & $2\omega - \Omega + N$\\
      \hline
      \(2^{nd}\) Orbital Harmonic  & 156.7026 \(\pm\) 0.0002 & 551.3 & \(2\omega\)\\
      \hline
      \(3^{rd}\) SH Harmonic  & 231.598\(\pm\) 0.003 &  373.0 & \(3\omega - 3\Omega\)\\
      \hline
      ?? & 232.765\(\pm\) 0.004 & 371.2 & \(3\omega - 2\Omega\)\\
      \hline
      ?? & 233.916\(\pm\) 0.004 &  369.3 & \(3\omega - \Omega\)\\
      \hline
      \(4^{th}\) SH Harmonic  & 308.803\(\pm\)  0.003 & 279.79 & \(4\omega - 4\Omega\)\\
      \hline
      ?? & 308.821\(\pm\)  0.005 & 279.77 & \(4\omega - 2\Omega\)\\
      \hline
      \(5^{th}\) SH Harmonic   & 385.994 \(\pm\) 0.004 &  223.8 &  \(5\omega - 5\Omega\)\\
      \hline
      \(6^{th}\) SH Harmonic  & 463.23\(\pm\) 0.02 &  186.5 & \(6\omega - 6\Omega\)\\
      \hline
      \(7^{th}\) SH Harmonic  & 540.43\(\pm\) 0.03&  159.9 & \(7\omega - 7\Omega\)\\
      \hline
      \(8^{th}\) SH Harmonic  & 617.63\(\pm\) 0.03&  139.9 & \(8\omega - 8\Omega\)\\
      \hline
      \(9^{th}\) SH Harmonic  & 694.84\(\pm\) 0.04&  124.3 &\(9\omega - 9\Omega\)\\

      \hline
     \end{tabular}
     \caption{Major frequencies detected in the lightcurve.}
     \centering
     \label{tab:frequencies}
    \end{table*}
    
        \begin{figure*}
        \includegraphics[width=20 cm, height=6 cm]{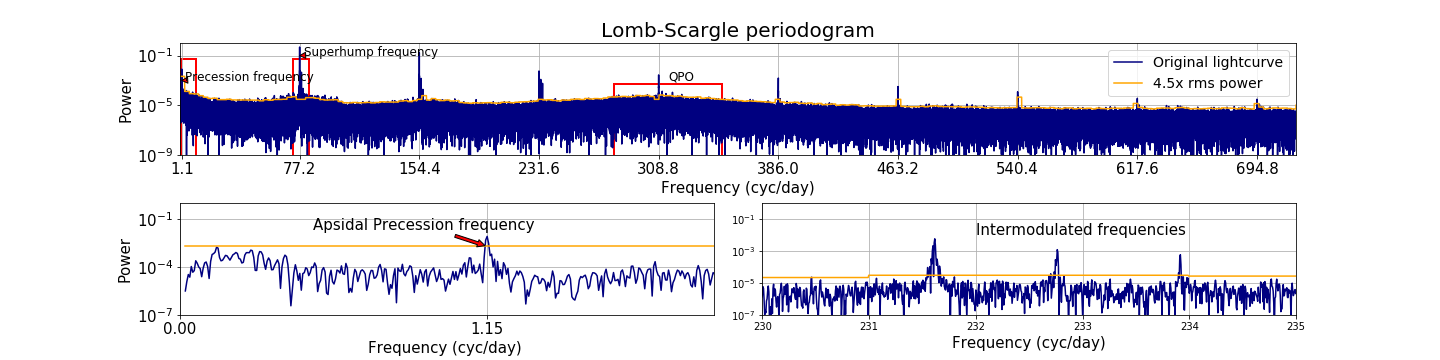}
        \caption{Lomb-Scargle periodogram of the lightcurve.  The yellow line indicates 4.5$\sigma$ significance above the noise.}
        
        \centering
        \label{fig:lombscargle}
    \end{figure*}
    
    \begin{figure*}
        \includegraphics[scale=1.2]{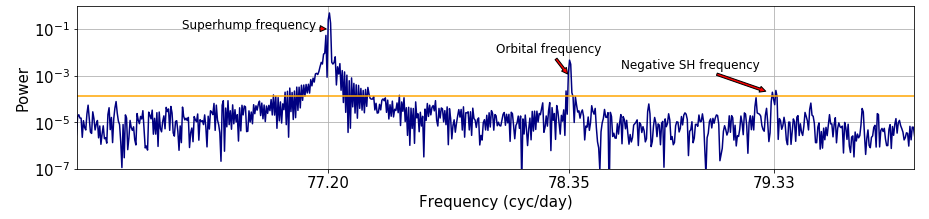}
        \caption{Lomb-Scargle periodogram around the superhump frequencies.}
        \centering
        \label{fig:lombscargle_sh}
    \end{figure*}

   
   Power spectra computed over discrete, independent time intervals for the four main frequencies are presented in Fig.~\ref{fig:sliding_fft}. The positive superhump and the orbital period are visible over the full observing period, though the orbital period is substantially weaker around 10 days and also at the end of the observing period.  The precession period is below the detection limit for weeks 2-3, but is well-detected at other times.  The potential negative superhump varies strongly, and is undetectable at quite a few epochs.
   
    The precession period is actually visible by eye in the lightcurve itself, if we average over the more rapid oscillations.  This is illustrated in Fig. \ref{fig:binnedlightcurve}, which shows a binned version of the lightcurve. Each point is taken to be 
    the flux average of 250 independent points from the original lightcurve. The binning duration of about 250 minutes per point is longer than both the superhump and the orbital periods and so they get averaged out in this binned lightcurve. The precession oscillations, with period of about 
    0.87 days, are clearly visible except in the 10 to 35 day window where the sliding power spectrum of Fig. \ref{fig:sliding_fft} also indicated that the oscillations were at low amplitude. This binned lightcurve also reveals aperiodic variability between days 10 and 55, for which we have 
    no explanation.
   
   Observed-minus-computed (O-C) diagrams are a powerful tool that can be used to search for period variations. These diagrams compare the timing of an event, which in our case is the time of phase zero in the lightcurve based on an ephemeris (observed), to when we expect such an event if it occurred at an exactly constant periodicity (computed). A linear trend in an O-C diagram corresponds to an incorrect period, whereas a parabolic trend is caused by a period derivative. To compute the O-C diagram, the lightcurve was broken into one day segments for the superhump. Each of these segments was folded onto the superhump frequency with a fixed zero-point and a quadratic polynomial was fitted near the peaks of these phase folded segments. The change in location of the maximum of the polynomial was taken as a measure of the phase shift of the oscillation. The time of maximum was then derived from this phase shift. A quadratic fit to different regions in the O-C diagram yielded a measurement of the period derivative \citep{2005ASPC..335....3S}.
    
    The O-C diagram for the positive superhump is shown in Fig.~\ref{fig:O-C}. Five parabolas were fit to different regions in the O-C diagram and the quadratic term gave an estimate of the rate of change of period $\dot{P}$. We observed period derivatives of \{$0.87$, $-1.61$, $2.02$, $-2.82$ and $0.47$\}$\times 10^{-7}$ days per day on these different segments.  By examining the light curve shape as a function of time, we have confirmed that the variation in the O-C diagram of the positive superhump is indeed due to these period changes and not, say, due to a change in the morphology of the lightcurve shape.  
    
    \begin{figure*}
        \centering
        \includegraphics[scale=0.35]{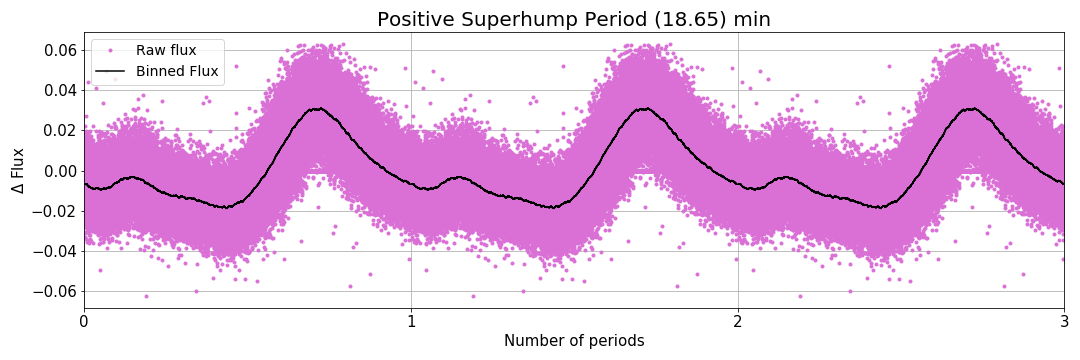}
        \includegraphics[scale=0.35]{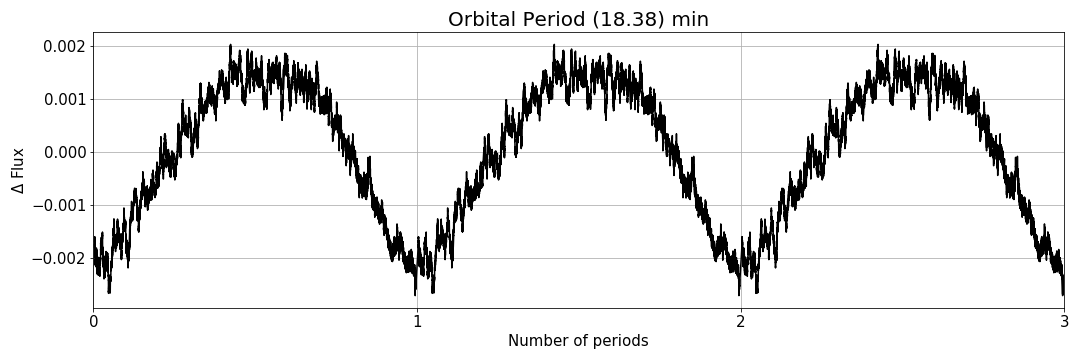}
        \includegraphics[scale=0.35]{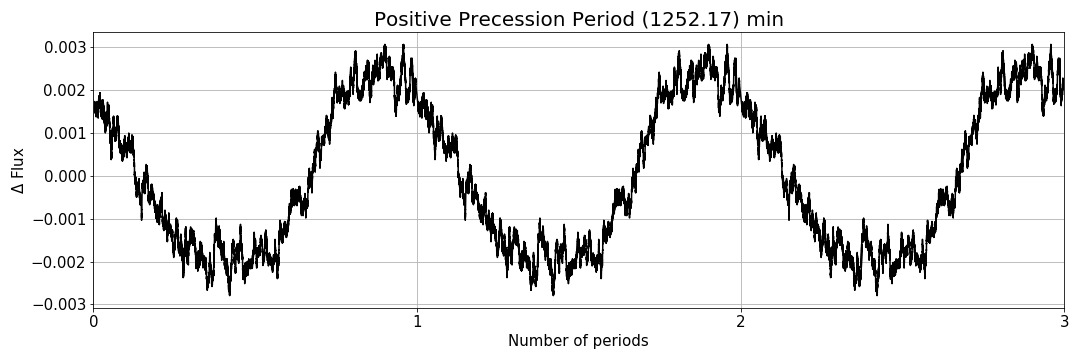}
        \includegraphics[scale=0.35]{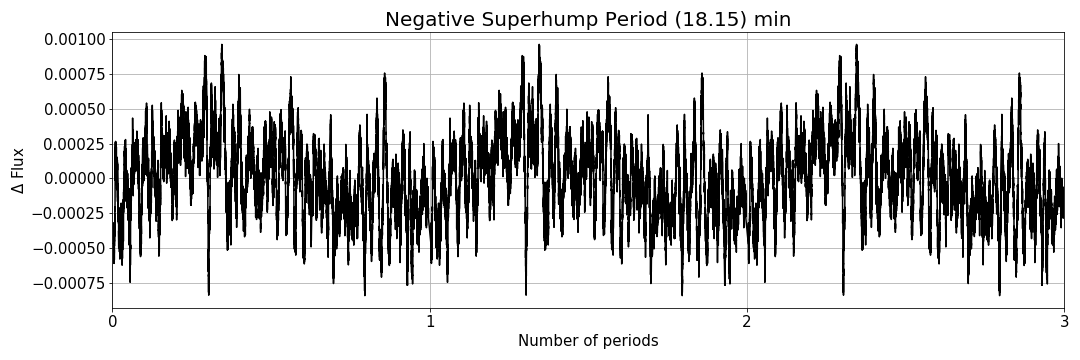}
        \caption{Phase folded and binned lightcurves on the positive superhump, orbital, positive apsidal precession and negative superhump periods respectively.  We also show the unbinned phase-folded lightcurve for the superhump in pink in the uppermost panel.}
        \label{fig:phase_fold}
    \end{figure*}

    \begin{figure*}
        \centering
        \includegraphics[scale=0.4]{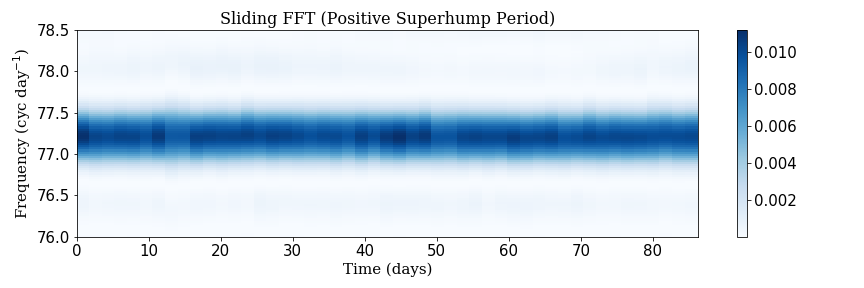}
        \includegraphics[scale=0.4]{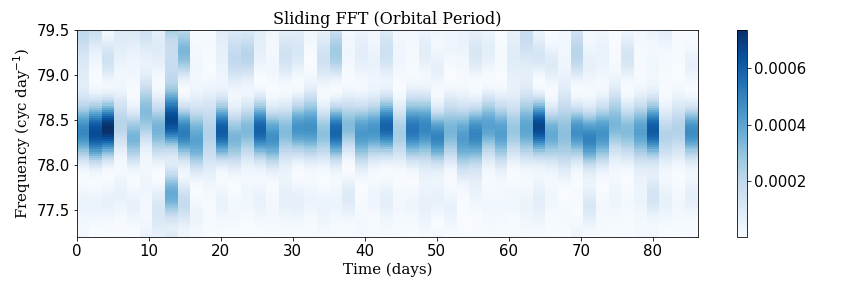}
        \includegraphics[scale=0.4]{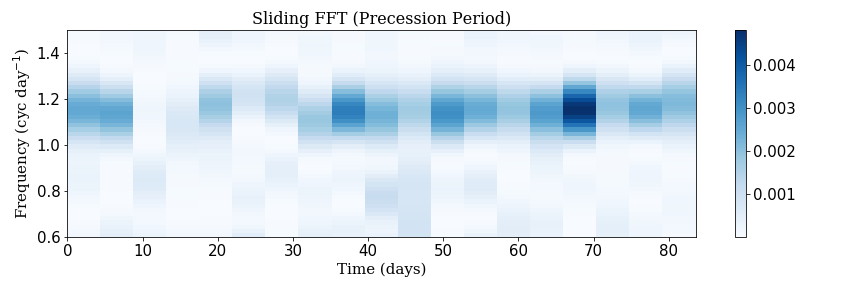}
        \includegraphics[scale=0.4]{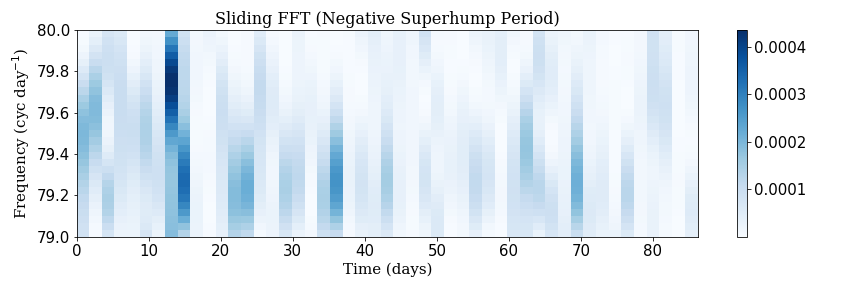}
        \caption{From top to bottom, the sliding power spectrum for the positive superhump, orbital, precession, and (potential) negative superhump variations, respectively, as a function of time since the start of the {\it K2} observing period. }
        \centering
        \label{fig:sliding_fft}
    \end{figure*}
    
        \begin{figure*}
        \centering
        \includegraphics[scale=1.5]{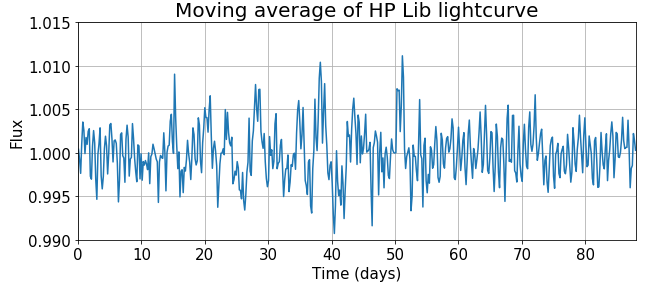}
        \caption{Entire 88 day lightcurve after binning on independent 250 min. intervals. Note that the superhump frequency is averaged out but the precession period is now visible as oscillations with a period of a little less than a day.}
        \centering
        \label{fig:binnedlightcurve}
    \end{figure*}
    
    \begin{figure*}
        \centering
        \includegraphics[scale=0.3]{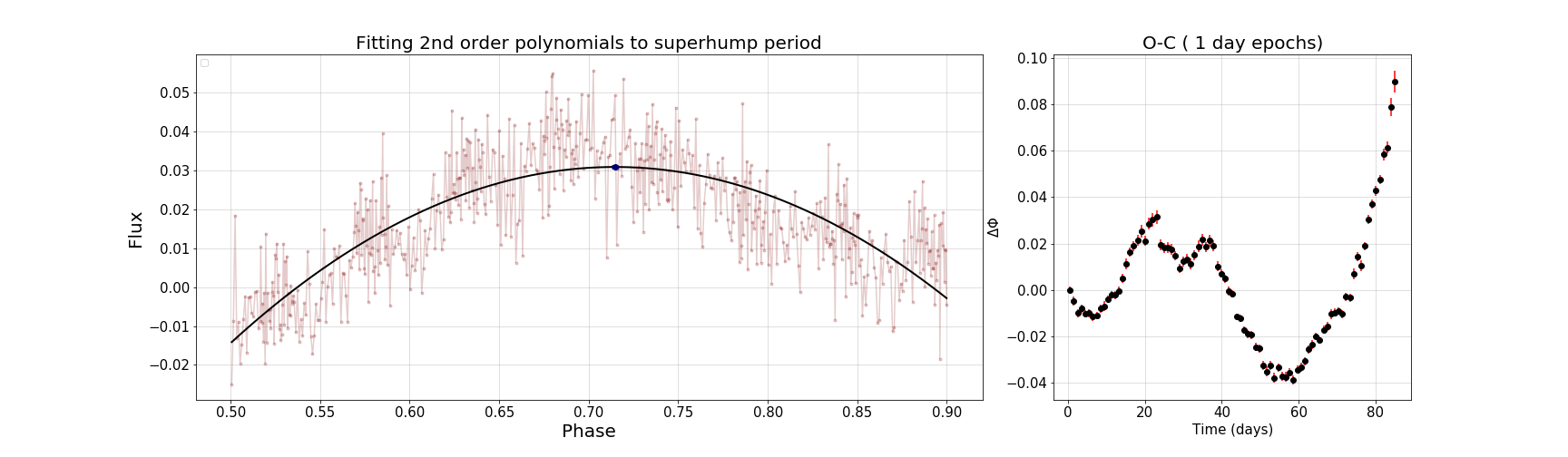}
        \caption{O-C diagram of the positive superhump period (right panel).  The left panel illustrates our fitting to the maximum in the superhump lightcurve over one of the epochs.}
        \centering
        \label{fig:O-C}
    \end{figure*}
    
While the focus of this paper is on discrete periodicities in the {\it K2} lightcurve, it is also worth commenting on the continuum noise power.  Fig.~\ref{fig:power-law} is a log-log plot of the power spectrum, computed this time using Welch's \footnote{https://docs.scipy.org/doc/scipy/reference/generated/
scipy.signal.welch.html} method.  This provides a better characterization of the high frequency continuum shape at the cost of losing information at low frequencies.  The discrete features we have been discussing are all present on this plot, including the precession frequency at 1.1501 cyc/day and the QPO at $\sim300$~cyc/day.  The excess power below 0.5 cyc/day may be due to the periodicity in the data gaps shown in Fig.~\ref{fig:lc_properties}.
We fit a power law in the frequency range 1.5 - 70~cyc/day, and found that the continuum is well fit by $P_\nu\propto\nu^{-0.58}$.  This power law fit successfully recovers the high frequency continuum power above the QPO, even though we did not fit it to this portion of the spectrum.  There is no evidence for a break in the continuum power over our detectable frequency range.

Fig.~\ref{fig:power-law} also shows that the QPO is centered on approximately four times
the superhump or orbital frequency, and that there are also lower power, broad QPO-like features roughly centered on the superhump/orbital frequency and twice that frequency.  We have fit all three of these QPOs with Lorentzian profiles and find that their peak frequencies and quality factors are 78.0~cyc/day ($Q=6.1$), 163.2~cyc/day ($Q=5.5$), and 305.7~cyc/day ($Q=3.5$).  Despite its lower quality factor, the $\sim300$~cyc/day QPO has 7-11 times more integrated power than the other QPOs.
    
     \begin{figure*}
        \centering
        \includegraphics[scale=0.4]{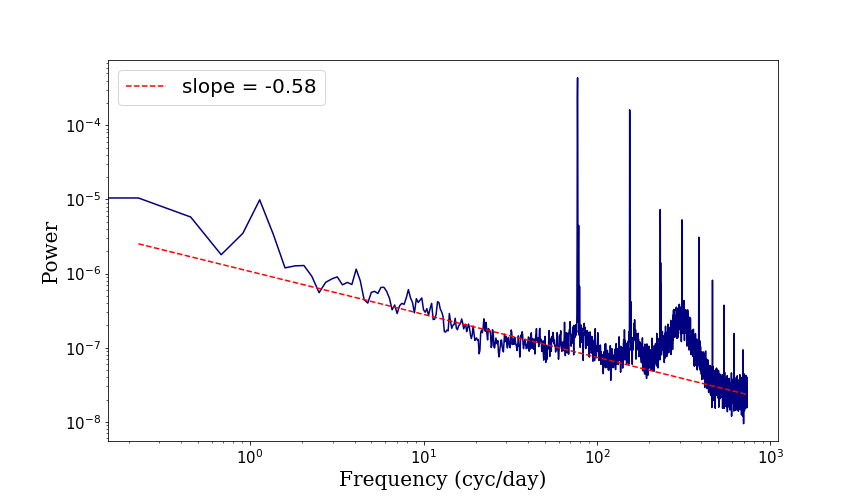}
        \caption{Power law fit to the continuum portion of the power spectrum.}  
        \centering
        \label{fig:power-law}
    \end{figure*}   

        

\section{Discussion}
    The phase-folded positive superhump lightcurve in
    Fig.~\ref{fig:phase_fold} is very similar in both amplitude and shape to the $V$-band lightcurve of HP~Lib observed by \citet{pat02}, consistent with the stability they reported in this feature over six years.  Those authors found variation in the much weaker $V$-band orbital period lightcurve between 1998 and 2001 (see Figure 4 from that paper).  Our phase-folded lightcurve better resembles the 1998 lightcurve in shape and amplitude ($\simeq0.003$ mag in 1998), though we do not appear to have the double-peaked shape near maximum that \citet{pat02} found.  Nevertheless, our orbital period light curve is flatter at maximum and sharper at minimum, which is qualitatively similar to what they found.  It is noteworthy that a somewhat flat-topped light curve near maximum with a sharper minimum was seen in the 1997 orbital period light curve of AM~CVn itself \citep{ski99}, similar to what we see here for HP Lib.  However, in 1998 the orbital period light curve was more double-peaked near maximum \citep{ski99}, similar to the earlier HP Lib observations of \citet{pat02}.  Hence these differences in orbital period lightcurve shape in HP Lib are similar to what has already been observed in AM~CVn.  The lightcurve shapes of the negative superhump in AM~CVn are consistent with being sinusoidal \citep{ski99}, similar to what we observe here in HP Lib for the lightcurve of the possible negative superhump.
    
    \citet{pat02} did not find any low frequency (0.1-5.0 cycles day$^{-1}$) variability in their six year, ground-based data of HP Lib down to an amplitude of 0.04~mag.  Our detection of a precession frequency of 1.14 cycles day$^{-1}$ with variation amplitude of $\simeq0.003$~mag is consistent with their upper limit.  The less than 0.001~mag amplitude of our putative negative superhump was well below their detection threshold.  Moreover, Fig.~\ref{fig:sliding_fft} shows that this feature is not always present and long term continuous data is required to detect the signal. \cite{pat02} also reported a QPO in the 280-350 cyc day\(^{-1}\) region, with maximum power near 310 cyc~day\(^{-1}\) which we also see in Figs.~\ref{fig:lombscargle} and \ref{fig:power-law}. The cause of power in this region is still not well understood, but it is remarkable how stable this feature is over a time range of more than a decade, given that most QPOs in, e.g., X-ray binaries show considerable variations in frequency (e.g. \citealt{don07}).
    
    
    The positive superhump frequency is given by the difference between the orbital
    ($\omega$) and apsidal precession ($\Omega$) frequencies: $\omega-\Omega$.
    We also searched for a corresponding (nodal?) precession period for the putative negative superhump and found a peak in the Lomb-Scargle periodogram that was not quite $3\sigma$ above the noise.  It might be possible in the future, with better observational data, to confirm this feature, which would shed light on the physical nature of the putative negative superhump frequency. On the other hand, it is interesting that we have detected one other intermodulation frequency involving the negative superhump, orbital, and apsidal precession frequencies.
    

    Even with the exquisite photometry of the {\it K2} data, we do not detect any discrete frequency features that cannot be connected to the basic binary orbital frequency and accretion disc precession frequencies in HP~Lib (assuming our identification of the negative superhump is correct).  This result is similar to most other high-state AM~CVn systems.  Ground-based photometry of AM~CVn itself
    revealed 19 discrete frequency features all of which could be related to the orbital frequency and the apsidal and nodal precession frequencies, though the latter two were not directly detected \citep{ski99}.  \cite{gre18} analyzed 80 days of {\it K2} data of the AM CVn, SDSS J1351, comparable to the data we present here but with a much lower cadence of 30 minutes.  They directly detected the apsidal precession period of 2.17 cycles/day, with a phase-folded lightcurve consistent with being sinusoidal.  They were also able to infer the orbital and positive superhump frequencies at 91.54 and 89.37 cyc/day, respectively, even though these were above their Nyquist frequency.  However, they confirmed these frequencies using high speed ground-based observations, and the phase-folded lightcurves of both the superhump and orbital period were similar to our superhump light curve.  They noted that this similarity may be the result of beating between the two signals.  However, the distinct phase-folded orbital period light variation that we find for HP~Lib is the same whether or not we subtract off the positive superhump variations first.
    
    There is one high-state AM~CVn system which is very different in its photometric variability:  SDSS~J1908.  \citet{fon11} first analyzed several months of {\it Kepler} photometric data on this source, and detected eleven discrete frequency features which they were able to show were integer combinations of three discrete frequencies, similar to other high-state AM~CVn systems including HP~Lib.  However, \citet{kup15} presented spectroscopy data showing that the actual orbital period
    (1085.1~s, just slightly shorter than that of HP~Lib:  1102.8~s) was {\it not} one of the periods that \citet{fon11} had detected.  Moreover, \citet{kup15} presented a full four years of {\it Kepler} photometry, and detected many more (42) discrete frequency peaks in the power spectrum, including 10 of the 11 that \cite{fon11} had found, and more than double the 18 we have found here for HP~Lib in Table~\ref{tab:frequencies}.  The orbital period was one of the photometric frequencies, and based in part on the morphology of the phase-folded lightcurve, \citet{kup15} tentatively identified an 1150.6~s periodicity (also not detected by \citet{fon11}) as the positive superhump.   The much larger
    number of frequencies detected by \citet{kup15} is perhaps in part
    due to their much longer {\it Kepler} light curve.  However, unlike here in HP~Lib,
    the positive superhump identified by \citet{kup15} is {\it not} the strongest signal in the power spectrum.  That honor belonged to a signal at 938.45~s and its harmonic at 469.22~s, both {\it smaller} than the orbital period.  These were detected by \citet{fon11}, and many of the new frequencies detected by \citet{kup15} are still linear combinations of the three frequencies identified in the original scheme of \citet{fon11}.  It is clear, though, that these are {\it not} the orbital and apsidal precession frequencies of the disc in this system.
    
    The putative positive superhump in SDSS~J1908 did show stronger photometric variations than those at the orbital period, in agreement with what we have found here for HP~Lib.  However, the inferred apsidal precession period of 19,060~s is much shorter than the 75,710~s that we have firmly detected in HP~Lib.  This would suggest a higher mass ratio in SDSS~J1908, given the usual correlation between fractional superhump minus orbital period difference and mass ratio \citep{pat05,kni06}.  SDSS~J1908 is unusual in other ways, too, as it has a much higher accretion rate $>10^{-7}$~M$_\odot$~yr$^{-1}$ for its orbital period compared to other AM CVn systems \citep{ram18}.  SDSS~J1908 may be a much younger system than the other high-state AM~CVn systems, possibly evolving toward shorter, rather than longer, orbital periods.\footnote{(\citet{kup15} attempted to measure the orbital period derivative using O-C techniques, but this was swamped by shorter time scale variability.)}  If so, the additional frequencies detected by \citet{kup15}, many of which still follow the original three-frequency combination scheme of \citet{fon11}, may be related to the relative youth of this system and/or the unusually high accretion rate.  Note that the high accretion rate places the white dwarf effective temperature well above the DB instability strip
    (Fig. 2 of \citealt{bil06}), suggesting that the extra frequencies may still originate in the accretion flow or boundary layer, rather than the white dwarf itself.

    SDSS~J1908 has a QPO at 200-230 cyc/day \citep{kup15}, a significantly lower frequency than the $\sim 310$ cyc/day QPO in HP~Lib given that these two systems have almost the same orbital periods.  QPOs have apparently not yet been detected in the other high-state AM~CVn systems.

    
    Using the empirical relation between the mass ratio $q=M_2/M_1$ and the superhump period excess measured in hydrogen accreting systems \citep{pat05}, the mass ratio for HP~Lib is approximately $q=0.074$, consistent with the earlier estimate of 0.07 by \citet{pat02}.  The tidal truncation radius $r_{\rm out}$ of the disk for this mass ratio is 0.55 times the binary separation $a$ \citep{whi91}.  As is well-known (e.g. \citealt{mur00}), the empirical relation requires a disk apsidal precession rate that is less than that of a test particle orbit at large radii.  One reason for this is that the precession is actually due to an average of the tidal torques over the entire disk.  An appropriate average based on the linear theory of eccentric waves \citep{goo06} is
    \begin{eqnarray}
        \Omega =\left[2\int_{r_{\rm in}}^{r_{\rm out}} r^3\Sigma(GM_1/r^3)^{1/2}|E|^2dr\right]^{-1}\cr
        \times\int_{r_{\rm in}}^{r_{\rm out}}\frac{\Sigma q r^5(GM_1/r^3)}{2a^2}b^{(1)}_{3/2}(r/a)|E|^2dr,
    \end{eqnarray}
    where $\Sigma(r)$ is the radial surface mass density profile of the disk, $E$ is the complex eccentricity, and $b^{(1)}_{3/2}$ is a Laplace coefficient.  We have neglected pressure effects here, which would act to slow the precession rate.  Assuming the surface density is proportional to $r^{-3/4}$ (the outer region of a standard \citealt{sha73} disk), neglecting the inner radius, and ignoring the radial dependence of $E$, we find $\Omega=0.0154$ times the orbital frequency, remarkably close to our observed value 0.0147 of this ratio.  This formula for the precession frequency scales with the outer disk radius as approximately $r_{\rm out}^{3/2}$.  Our observed superhump period derivatives correspond to approximately one percent changes in the period over our observed time scales, and this can therefore be explained by approximately one percent changes in the outer disk radius over these time scales.

\section{Conclusions}
    In this paper we analyzed the {\it Kepler/K2} lightcurve of the AM CVn system HP Lib. 
    We observed four primary discrete frequencies which we identified as the positive superhump, 
    orbital, positive apsidal precession and (possibly) negative superhump frequencies in decreasing order of Lomb-Scargle power, respectively.  This marks only the second time that an apsidal precession frequency has been photometrically detected in an AM~CVn system.  We also detected 14 other discrete frequencies which were linear combinations of these four, in addition to the positive superhump itself which is a linear combination of the orbital and apsidal precession frequencies.  Like many well-observed AM~CVn systems, with the notable exception of SDSS~J1908 \citep{kup15}, HP~Lib is a very clean system, with discrete variability that can all be tied to just a few frequencies.
    
    We presented phase-folded lightcurves of the variability at the four primary frequencies, and found the shapes to be consistent with those observed in many other systems.  Using O-C techniques, we were able to measure period changes in the positive superhump frequency $\simeq10^{-7}$ days per day, consistent with one percent variations in the outer disk radius over the 88 day observing period.
    
    Finally, we confirmed the existence of a QPO at $\sim300$ cyc/day, a feature which is apparently stable over decades of observing.  This QPO is actually the highest frequency member of a set of three QPOs centered on the superhump/orbital frequency and its second and fourth harmonics.  This harmonic relation is suggestive of some sort of wave origin associated with the eccentric disk.  The continuum power spectrum is consistent with a single power law $P_\nu\propto\nu^{-0.58}$, extending both below and above the QPOs with no evidence of any breaks.

\section*{Acknowledgements}

We are grateful to the referee, Phil Charles, for a thoughtful report that led to additional insights.  We thank Matthew Middleton, Bryance Oyang, and Chris White for useful conversations, and Samuel Lu and Xin Sheng for identifying some of the periodicities in SDSS~J1908 within the three frequency scheme of \citet{fon11}.  This work was supported in part by the US National Science Foundation through grants ACI-1663688 and PHY-1748958, and by NASA through grant 80NSSC18K0727.  EB acknowledges support from the UK Science and Technology Facilities Council, grant number ST/S000623/1. SS acknowledges support from NASA under grant number 80NSSC20K0439.

\section*{Data Availability}

All the observational data on which the results of this paper are based are publicly available at the Kepler Data Archive.  The results of our analysis of this data, in particular the extracted lightcurve, can be obtained from the authors upon request.


\bibliographystyle{mnras}
\bibliography{refs}




%

\bsp	
\label{lastpage}
\end{document}